\documentclass[aps,prd,preprint,showpacs,nofootinbib,showkeys]{revtex4}
\usepackage[dvips]{epsfig}
\usepackage{amsfonts,amssymb}
\usepackage{indentfirst}
\usepackage{subfigure}

\newcommand{\la}{\zeta}
\newcommand{\lap}{\zeta_+}
\newcommand{\lam}{\zeta_-}
\newcommand{\lapm}{\zeta_\pm}

\newcommand{\J}{\ensuremath{\mathcal{S}}}
\newcommand{\E}{\ensuremath{\mathcal{E}}}
\newcommand{\xim}{\ensuremath{\xi_{\mathrm{max}}}}

\newcommand{\be}{\begin{equation}}
\newcommand{\ee}{\end{equation}}
\newcommand{\ben}{\begin{displaymath}}
\newcommand{\een}{\end{displaymath}}
\newcommand{\bea}{\begin{eqnarray}}
\newcommand{\eea}{\end{eqnarray}}\newcommand{\bean}{\begin{eqnarray*}}
\newcommand{\eean}{\end{eqnarray*}}

\newcommand{\ads}[1]{\mbox{${AdS}_{#1}$}}

\newcommand{\eg}{{\it e.g.}}
\newcommand{\ie}{{\it i.e.}}

\newcommand{\commentout}[1]{}

\newcommand{\beq}{\begin{equation}}
\newcommand{\eeq}{\end{equation}}
\newcommand{\beqr}{\begin{displaymath}}
\newcommand{\eeqr}{\end{displaymath}}
\newcommand{\beqa}{\begin{eqnarray}}
\newcommand{\eeqa}{\end{eqnarray}}
\newcommand{\beqar}{\begin{eqnarray*}}
\newcommand{\eeqar}{\end{eqnarray*}}
\renewcommand{\r}{\rho}
\newcommand{\cH}{{\cal H}}
\newcommand{\cN}{{\cal N}}

\newcommand{\half}{\ensuremath{\frac{1}{2}}}

\newcommand{\N}[1]{\ensuremath{\cN=#1}}

\newcommand{\Deltaxi}{\ensuremath{\Delta \xi}}

\begin{document}


\title{\Large Double--helix Wilson loops: case of two angular momenta}

\author{Andrew Irrgang}
\email{irrgang@purdue.edu} 
\author{Martin Kruczenski}
\email{markru@purdue.edu}
\affiliation{Department of Physics, Purdue University, 525 Northwestern Avenue,
        W. Lafayette, IN 47907-2036.}

\date{\today}

\begin{abstract}
Recently, Wilson loops with the shape of a double helix have played an important role in studying 
large spin operators in gauge theories. They correspond to a quark and an anti-quark moving in circles
on an $S^3$ (and therefore each of them describes a helix in $R\times S^3$). In this paper we consider
the case where the particles have two angular momenta on the $S^3$. The string solution corresponding 
to such Wilson loop can be found using the relation to the Neumann-Rosochatius system allowing the computation
of the energy and angular momenta of the configuration. The particular case of only one
angular momentum is also considered. It can be thought as an analytic continuation of the rotating strings
which are dual to operators in the SL(2) sector of \N{4} SYM.
\end{abstract}

 



\maketitle

\section{Introduction}
\label{intro}

 Wilson loops play an important role in understanding the AdS/CFT correspondence \cite{malda}. In the dual string picture they can be computed 
by finding a minimal area surface \cite{WL1} which has lead to various interesting prediction for their gauge theory counterparts \cite{WL2}. They can also be 
used to compute other quantities, for example the anomalous dimension of twist two operators for large spin \cite{LLWL} and scattering amplitudes \cite{SAWL}.

 The relation of Wilson loops to twist two operators appears through the cusp anomaly \cite{LLWL} but also in a different way that we exploit here.  
In fact twist two operators are described by the folded rotating string in AdS studied by Gubser, Klebanov and Polyakov \cite{GKP}.
When the spin grows to infinity the string touches the boundary and, in the field theory side, it can be replaced by a light-like Wilson loop.
In the $S^3\times R$ boundary this Wilson loop has the shape of a double helix since it describes two particles rotating in circles. 
 This point of view was emphasized recently in \cite{AM}. On the other hand, such double-helix Wilson loop can also be seen as the limit of a 
Wilson loop in which the particles move in circles with velocity $v<1$. When $v\rightarrow 1$ we again recover the same configuration. Motivated by 
this we consider here Wilson loops describing a quark and an anti-quark moving on an $S^3$ such that the angular separation between 
them remains fixed. The corresponding picture in the string side is a hanging string moving with two possible angular momenta in $\ads{5}$. 
It can be obtained by an ansatz similar to \cite{spikes} which reduces the problem to a version of the Neumann-Rosochatius integrable system \cite{NR}. 
This is a generalization of the ansatz proposed by Drukker and Fiol in \cite{DF}. In fact, in \cite{DF} already certain Wilson loops with the shape 
of a double helix in Euclidean space were described \footnote{They were actually called ``double-helix'' in the later work \cite{DGGM} where certain 
related Wilson loops were also studied.}. For the solutions obtained we compute the energy $E$ and the angular momenta $J_{1,2}$.  It should be noted 
that, as for most Wilson loops, the energy diverges near the boundary so it should be regularized by subtracting a term that can be interpreted as the 
self-energy of the quark (or anti-quark). Since, in this case, the quark 
is moving, the infinite self-energy gives rise to an infinite contribution to the angular momenta that should also be subtracted. In the case where one angular momentum vanishes, 
the result seems as an analytic continuation of the spiky string related to higher twist operators. This is intriguing since higher twist operators, in the $SL(2)$ sector have 
a description in terms of spin chains \cite{mz}. It would be interesting to find a similar description for the Wilson loops. In fact, for open string ending on D-branes inside 
AdS such description is already known \cite{OSSC} but as far as we know not for this case. The paper is organized as follows. In the next section we consider the simple case
of one angular momentum where everything can be computed explicitly in terms of elliptic functions. In the following section we study the case of two angular momenta. Finally
we consider a special case that has to be treated separately and later give our conclusions. The appendices are devoted to certain parts of the calculation including an initial 
approach to the field theory side where we compute the electromagnetic field of two charges of opposite sign moving in circles in a three-sphere.

\section{One angular momentum} \label{sec:1S}

 The simplest case one can consider is when the two particles are moving on a maximum circle of the $S^3$ in which case the
string has only one non-vanishing angular momentum.  The boundary metric is
\beq
 ds^2 = -dt^2 + d\theta^2 + \sin^2\!\theta \, d\phi_1^2 + \cos^2\!\theta \, d\phi_2^2
\eeq
where $\theta$, $\phi_1$ and $\phi_2$ parameterize an $S^3$. The quark and anti-quark are located at 
$\theta=\phi_1=0$ and $\phi_2=\pm \half \Delta\phi +\omega t$, where $\omega$ is the angular velocity. On the string side, the metric is
\beq
ds^2 = -\cosh^2\!\rho\, dt^2 + d\rho^2 + \sinh^2\!\rho\,  \left( d\theta^2 + \sin^2\!\theta \, d\phi_1^2 + \cos^2\!\theta \, d\phi_2^2 \right)
\label{S3m}
\eeq
 The corresponding string ends in the boundary at the location of the quark and anti-quark. We can parameterize the string by $(\sigma,\tau)$ as
\beq
 t =\tau, \ \ \ \phi_2=\sigma+\omega \tau, \ \ \ \rho=\rho(\sigma,\tau) = \rho(\sigma), \ \ \ \theta=\phi_1=0
\eeq
 where for the coordinates $t$ and $\phi_2$ we made a gauge choice (static gauge) and for the others we consider a particular ansatz. In fact 
it can be seen that such an ansatz solves all the equations of motion if
\beq
\rho'(\sigma) = \half \frac{\sinh2\rho}{\sinh2\rho_0} \sqrt{\frac{\sinh^2 2\rho - \sinh^2 2\rho_0}{\cosh^2 \rho - \omega^2 \sinh^2\rho }}
\eeq
is satisfied. Here $\rho_0$ is a constant of integration. In fact the equation is the same as in \cite{spikes}, we are only interested in a 
different solution. The energy and spin can be computed to be
\beqa
S &=& \omega \int_{\rho_0}^{\rho_M} \frac{\sinh \rho}{\cosh\rho} \sqrt{\frac{\sinh^2 2\rho - \sinh^2 2\rho_0}{\cosh^2 \rho - \omega^2 \sinh^2\rho }} \\
E -\omega S &=& 2 \int_{\rho_0}^{\rho_M} d\rho \sinh2\rho \, \sqrt{\frac{\cosh^2 \rho - \omega^2 \sinh^2\rho}{\sinh^2 2\rho - \sinh^2 2\rho_0}} .
\eeqa
 Here the integrals diverge near the boundary so we introduce a cut-off $\rho_M\gg 1$. Other than that they are the same integrals as for the rotating string
\cite{spikes} extrapolated from $\omega>1$ to $\omega<1$ and in that sense we say that these formulas are an analytic continuation of the rotating string. 
As in that case, the integrals can be computed in terms of elliptic integrals giving
\beqa
S &=& \sqrt{x_0-1}\, e^{\rho_M} \nonumber \\
   && + 2 \frac{\sqrt{x_1+x_0}}{\sqrt{x_0-1}}  \left\{ x_1  K(q) - (x_0-1) E(q) 
      - \frac{x_1(1+x_1)}{x_1+x_0} \Pi\left(\frac{x_0-1}{x_0+x_1},q\right) 
     \right\}  \label{Sres} \\
E -\omega S &=& \frac{1}{\sqrt{x_0}} e^{\rho_M} + \frac{2\sqrt{x_0+x_1}}{\sqrt{x_0}} \left( K(q) - E(q) \right)  \label{Eres}
\eeqa
where $x_0=\frac{1}{1-v^2}$, $x_1=\sinh^2\!\rho_0$, $q=\sqrt{\frac{x_0-x_1-1}{x_0+x_1}}$ and we kept only the terms which do not 
vanish when $\rho_M\rightarrow\infty$. 
To extract the divergent piece it is useful to note that
\beq
\Pi\left(\frac{\pi}{2}-\epsilon,1,q\right) 
  = \frac{1}{\sqrt{1-q^2}} \frac{1}{\epsilon} + K(q) - \frac{1}{1-q^2} E(q) + \ldots, \ \ \ \ \ \epsilon\rightarrow 0
\eeq
where the terms omitted vanish when $\epsilon\rightarrow 0$. To obtain a result finite in the limit $\rho_M\rightarrow\infty$ we subtract
the energy and momentum of a straight string ending on the boundary and moving with speed $v$. In the appendix we do such a computation to
show that the result is exactly equal to the divergent piece of the result found here. Therefore the difference is given by the finite piece
of the expressions (\ref{Sres}), (\ref{Eres}):
\beqa
\bar{S} &=&  2 \frac{\sqrt{x_1+x_0}}{\sqrt{x_0-1}}  \left\{ x_1  K(q) - (x_0-1) E(q) 
      - \frac{x_1(1+x_1)}{x_1+x_0} \Pi\left(\frac{x_0-1}{x_0+x_1},q\right) 
     \right\}  \label{Sfin} \\
\bar{E} -\omega \bar{S} &=&  \frac{2\sqrt{x_0+x_1}}{\sqrt{x_0}} \left( K(q) - E(q) \right)  \label{Efin}
\eeqa 
where the bars represent the finite values of $E$ and $S$, namely after subtracting the reference configuration.
One interesting limit that can be studied is when $v\rightarrow 1$ which results in 
\beqa
 \bar{S} &\simeq& -2x_0 - \half \ln x_0 -2 \ln 2 +\frac{3}{2} +\half \ln(1+2x_1) -2\sqrt{x_1(1+x_1)}\arctan\sqrt{\frac{x_1}{1+x_1}} \nonumber \\
 \bar{E}-\bar{S} &\simeq& \ln x_0 -\ln(1+2x_1)+4\ln2 -1 
\eeqa
The result can also be written as:
\beq
 \bar{E} \simeq \bar{S} + \frac{\sqrt{\lambda}}{2\pi} \ln |\bar{S}| + \ldots
\eeq
where we restored the tension of the string $T=\frac{\sqrt{\lambda}}{2\pi}$. Notice that in this case $\bar{E}<0$ and $\bar{S}<0$. This is because 
$\bar{E}$ and $\bar{S}$ represent the difference between  the actual $E$ and $S$ (which are positive) and those of a straight string. 
In fact $\bar{E}<0$ means that the force between the quark and anti-quark is attractive \cite{WL1}.
 The coefficient of $\ln S$ is the cusp anomaly (it is half the usual value because we consider an open string). The reason is that the limiting shape is the same as that 
of the spiky string when $\omega\rightarrow 1$ \cite{spikes}.

\section{Generalized NR ansatz for strings ending on the boundary}

We are now interested in strings moving in the full \ads{5} which can be defined as a subspace of $\mathbb{R}^6$ parameterized by
three complex coordinates, $X_{a=1,2,3}$, subject to the constraint
\begin{equation}
-1 =|X_1|^2 + |X_2|^2-|X_3|^3 = \sum_a \eta_a X_a \bar{X_a}.
\label{constraint}
\end{equation}
where, for brevity, we defined $\eta_a$ to be
\beq
\eta_1 = \eta_2 =1, \ \  \eta_3 = -1.      
\eeq
The metric is
\begin{equation}
ds^2 = \sum_a \eta_a dX_a d \bar{X}_a.
\end{equation}
It is also convenient to use the coordinates:
\beq
X_1 = \sinh \rho \sin\theta \, e^{i\phi_1}, \ \  X_2 = \sinh \rho \cos\theta \, e^{i\phi_2}, \ \  X_3 = \cosh \rho  \, e^{it}, 
\label{global}
\eeq
in which case the metric is
\beq
ds^2 = -\cosh^2\rho \, dt^2 + d\rho^2 + \sinh^2\rho\left(d\theta^2+\sin^2\theta d\phi_1^2 +\cos^2\theta d\phi_2^2\right) .
\eeq
namely (\ref{S3m}). Now however we put angular momentum in both $\phi_1$ and $\phi_2$. 
Going back to the coordinates $X_a$, the Lagrangian, in conformal gauge, is 
\begin{equation}
\mathcal{L} = \frac{T}{2} \sum_a \left[\eta_a \partial_\tau X_a \partial_\tau \bar{X}_a - \eta_a \partial_\sigma X_a \partial_\sigma \bar{X}_a \right] 
- \frac{T \Lambda}{2} \left(1+\sum_a \eta_a X_a \bar{X}_a\right) , \label{CGL}
\end{equation}
where $\Lambda$ is a Lagrange multiplier and $T$ is the tension of the string. The equations of motion for $X_a$ are
\beq
-\partial^2_\tau X_a + \partial_\sigma^2 X_a - \Lambda X_a = 0     ,                  \label{EOM}  
\eeq
and the constraints are
\begin{eqnarray}
\sum_a \eta_a\left[|\partial_\tau X_a|^2 + |\partial_\sigma X_a|^2\right] = 0                                             \label{CC1} \\
\sum_a \eta_a\left[\partial_\tau X_a \partial_\sigma \bar{X}_a + \partial_\tau \bar{X}_a \partial_\sigma X_a \right]= 0.  \label{CC2}
\end{eqnarray}
Since the tension $T$ drops out of the equations of motion we can temporarily ignore it and restore it at the end when we compute the
conserved quantities, \ie\ energy and angular momentum.

\subsection{Generalized Neumann - Rosochatius (NR) Ansatz}

The problem can be solved by reducing it to a hyperbolic version of the integrable Neumann-Rosochatius system \cite{NR}.  
We define $\xi\equiv\alpha\sigma+\beta \tau$ and propose the ansatz  
\beq
X_a       = x_a(\xi)\ e^{i\omega_a\tau}          \label{NRA} 
\eeq
to satisfy the equations of motion, (\ref{EOM}), and the conformal constraints, (\ref{CC1}) and (\ref{CC2}).  
Solutions of this form describe a rigid rotating string characterized by three frequencies, $\omega_{a=1,2,3}$, and 
whose shape is given by $x_a(\xi)$. 
The equations of motion for $x_a(\xi)$ and $\bar{x}_a(\xi)$ are determined by substitution into (\ref{EOM})
resulting in 
\beq
(\alpha^2-\beta^2) x''_a - 2 i \beta \omega_a x'_a +\omega_a^2 x_a - \Lambda x_a = 0                          \label{eom} 
\eeq
The conformal constraints in terms of $x_a$ and $\bar{x}_a$ are
\begin{eqnarray}
\sum_a \eta_a \left[2 \beta x'_a \bar{x}'_a - i\omega_a(x'_a \bar{x}_a - \bar{x}'_a x_a)  \right] &=& 0  \nonumber \\ 
\sum_a \eta_a \left[(\alpha^2+\beta^2)x'_a \bar{x}'_a - \beta\omega_a i(x'_a \bar{x}_a - \bar{x}'_a x_a) + \omega_a^2 x_a \bar{x}_a\right] &=& 0. 
\end{eqnarray}
or, equivalently, 
\begin{eqnarray}
\sum_a \eta_a [(\alpha^2-\beta^2)x'_a \bar{x}'_a + \omega_a^2 x_a \bar{x}_a] &=& 0                                    \label{cc1}\\
\sum_a \eta_a [i(\alpha^2-\beta^2)\omega_a(x'_a \bar{x}_a - \bar{x}'_a x_a) +2\beta \omega_a^2 x_a \bar{x}_a] &=& 0.  \label{cc2} 
\end{eqnarray}
Note that the equations of motion, (\ref{eom}), can be thought as following from the Lagrangian:
\beq
\mathcal{L} = \sum_a\eta_a\left[(\alpha^2-\beta^2) x'_a \bar{x}'_a + i\beta\omega_a(x'_a\bar{x}_a - \bar{x}'_a x_a) 
    - \omega_a^2 x_a \bar{x}_a\right] + \Lambda[ 1 + \sum_a \eta_a x_a \bar{x}_a]. 
\label{Lag}
\eeq
This Lagrangian determines the shape of the string. However, we can equivalently think of it as describing the motion of a particle with 
$\xi$ interpreted as time. Using this mechanical analogy we define the momenta canonically conjugate to $x_a$ and $\bar{x}_a$,  
\begin{eqnarray}
p_a       &= \frac{\partial \mathcal{L}}{\partial \bar{x}'_a} =& \eta_a\left[(\alpha^2 - \beta^2)x'_a       - i\beta\omega_a x_a\right], \\
\bar{p_a} &= \frac{\partial \mathcal{L}}{\partial x'_a}       =& \eta_a\left[(\alpha^2 - \beta^2)\bar{x}'_a + i\beta\omega_a \bar{x}_a\right] ,
\end{eqnarray}
and the Hamiltonian,
\begin{equation}
\mathcal{H} = \frac{1}{\alpha^2-\beta^2} 
  \sum_a \eta_a\left[p_a\bar{p}_a + i\beta\omega_a\eta_a(x_a\bar{p}_a-\bar{x}_a p_a)+\omega_a^2 x_a \bar{x}_a\right]. 
\label{Ham}
\end{equation}
It is now convenient to do a further change from the complex variables $x_a$ to real functions, $r_a(\xi)$ and $\mu_a(\xi)$,
\begin{equation}
x_a(\xi) = r_a(\xi) \, e^{i\mu_a(\xi)}.
\end{equation}
Namely the solution has the general form
\beq
X_a(\xi) = r_a(\xi) \, e^{i(\mu_a(\xi) + \omega_a \tau)} . 
\eeq
The hyperbolic constraint, (\ref{constraint}), becomes 
\begin{equation}
\sum_a \eta_a r_a^2 = -1. \label{rcon}
\end{equation}
The Lagrangian in terms of $r_a(\xi)$ and $\mu_a(\xi)$ is\footnote{We rescale the Lagrangian by a factor of two for convenience.}
\beq
\mathcal{L} = \half \sum_a \eta_a\bigg[(\alpha^2-\beta^2){r'}_a^2 + (\alpha^2-\beta^2)r_a^2 
   \bigg(\mu'_a - \frac{\beta\omega_a}{\alpha^2-\beta^2}\bigg)^2 -  \frac{\alpha^2\omega_a^2 r_a^2}{\alpha^2-\beta^2}\bigg] 
+ \half \Lambda \big(1+\sum_a \eta_a r_a^2\big). \label{rLag}
\eeq   
The momenta canonically conjugate to $r_a$ and $\mu_a$ are
\begin{eqnarray}
P_a = \frac{\partial \mathcal{L}}{\partial r'_a} &=& \eta_a(\alpha^2 - \beta^2)r'_a                                                                \\
C_a = \frac{\partial \mathcal{L}}{\partial \mu'_a} &=& \eta_a(\alpha^2 - \beta^2)r_a^2\bigg(\mu'_a - \frac{\beta\omega_a}{\alpha^2-\beta^2}\bigg). 
\end{eqnarray}
and the Hamiltonian,
\begin{eqnarray}
\mathcal{H} 
  = \half \frac{1}{(\alpha^2-\beta^2)}\sum_a \bigg[\eta_aP_a^2+\frac{\eta_aC_a^2}{r_a^2}+2\beta\omega_aC_a+\alpha^2\eta_a\omega_a^2r_a^2 \bigg]. 
\label{rHam}
\end{eqnarray}
The momenta $C_a$ are conserved, namely independent of $\xi$, which gives
\begin{equation}
\mu'_a = \frac{1}{\alpha^2-\beta^2} \left[\frac{\eta_aC_a}{r_a^2}  + \beta \omega_a \right].
\end{equation}
Using this, the equation of motion for the $r_a$ can be written as
\beq
(\alpha^2-\beta^2)r''_a - \frac{C_a^2}{(\alpha^2-\beta^2)r_a^3} + \frac{\alpha^2\omega_a^2 r_a}{\alpha^2-\beta^2}   - \Lambda r_a = 0. \label{eom_r}
\eeq
Using the conservation of the $C_a$'s, we obtain that the first constraint implies that the Hamiltonian vanishes:
\begin{equation}
\mathcal{H}
 =\half \sum_a \eta_a\bigg[(\alpha^2-\beta^2){r'}_a^2+\frac{1}{\alpha^2-\beta^2} \frac{C_a^2}{r_a^2}
      +\frac{\alpha^2\omega_a^2r_a^2}{\alpha^2-\beta^2}\bigg] + \frac{\beta}{\alpha^2-\beta^2} \sum_a \omega_a C_a= 0 
\label{rcc1}
\end{equation}
whereas the second constraint (\ref{cc2}) becomes
\begin{equation}
\sum_a \omega_a C_a = 0. \label{rcc2}
\end{equation}

\subsection{Unconstrained variables}

In order to solve the radial equations of motion for the case of two non-zero angular momenta, we use 
two unconstrained variables $\lapm$ related to $r_a$ through \cite{NR}
\begin{equation}
\sum_{a=1}^3 \frac{\eta_a r_a^2}{\la-\omega_a^2} = -\frac{(\la-\lap)(\la-\lam)}{\prod_{a=1}^3 (\la-\omega_a^2)}. 
\end{equation}
or equivalently 
\begin{eqnarray}
r_1^2 + r_2^2 - r_3^2                                        &=& -1           \\ \nonumber 
\sum_a \omega_a^2 + \sum_a \eta_a \omega_a^2 r_a^2           &=& \lap+\lam    \\ 
-\left(\prod_a\omega_a^2\right)\, \times \, \sum_b \eta_b\frac{r_b^2}{\omega_b^2} &=& \lap\lam.    
\end{eqnarray}
 More explicitly
\begin{equation}
 r_a^2 = -\eta_a\frac{(\lap-\omega_a^2)(\lam-\omega_a^2)}{\prod_{b\neq a} (\omega_a^2-\omega_b^2)}. \label{r(zeta)}
\end{equation}
The Lagrangian can be rewritten as
\begin{eqnarray}
\mathcal{L}&=&  \frac{(\alpha^2-\beta^2)(\lap-\lam)}{4} \left( \frac{{\lap'}^2}{\prod_a(\lap-\omega_a^2)} 
   - \frac{{\lam'}^2}{\prod_a(\lam-\omega_a^2)} \right) \nonumber \\ 
&& - \frac{1}{(\alpha^2-\beta^2)}\frac{1}{(\lap-\lam)}  \left( \sum_a \prod_{b\neq a} (\omega_a^2-\omega_b^2) 
   \left[\frac{C_a^2}{\lap-\omega_a^2}-\frac{C_a^2}{\lam-\omega_a^2}\right]\right) \nonumber  \\
&&  +\frac{\alpha^2}{\alpha^2-\beta^2} \left( \sum_a\omega_a^2-(\lap+\lam)\right). \label{ztLag}
\end{eqnarray}
The momenta canonically conjugate to $\lapm$ are defined to be
\begin{eqnarray}
p_+ = \frac{\partial \mathcal{L}}{\partial \lap'} &=&  \frac{(\alpha^2-\beta^2)(\lap-\lam)}{2\prod_a(\lap-\omega_a^2)}  \\
p_- = \frac{\partial \mathcal{L}}{\partial \lam'} &=& -\frac{(\alpha^2-\beta^2)(\lap-\lam)}{2\prod_a(\lap-\omega_a^2)},
\end{eqnarray}
and then the Hamiltonian is
\begin{eqnarray}
\mathcal{H} = \frac{1}{(\alpha^2-\beta^2)(\lap-\lam)}\bigg[\prod_a(\lap-\omega_a^2)p_+^2-\prod_a(\lam-\omega_a^2)p_-^2\bigg]
	\nonumber \\ 
+ \sum_a\frac{\prod_{b\neq a}(\omega_a^2-\omega_b^2)}{(\alpha^2-\beta^2)(\lap-\lam)}\bigg[\frac{C_a^2}{\lap-\omega_a^2}
           -\frac{C_a^2}{\lam-\omega_a^2}\bigg] \nonumber \\ 
- \frac{\alpha^2}{\alpha^2-\beta^2}\bigg[\sum_a \omega_a^2 - (\lap-\lam)\bigg].
\end{eqnarray}

\subsection{Solving the System with the Hamilton-Jacobi Method}

The Hamiltonian can be written in a more suggestive form by defining $\tilde{H}(p,\la)$ such that
\begin{eqnarray}
\tilde{H}(p,\la) &=& \prod_a(\la-\omega_a^2)\ p^2 + \sum_a C_a^2 \frac{\prod_{b\neq a} (\omega_a^2-\omega_b^2)}{\la-\omega_a^2} 
                        - \alpha^2 \sum_a\omega_a^2\ \la + \alpha^2 \la^2 \\
\mathcal{H} &=& \frac{1}{(\alpha^2-\beta^2)(\lap-\lam)} \left\{\tilde{H}(p_+,\lap) - \tilde{H}(p_-,\lam)\right\}. \label{ztHam}
\end{eqnarray}
The Hamilton-Jacobi method requires finding a function $\mathcal{W}(\lap,\lam)$ such that
\begin{equation}
\mathcal{H}\left(p_\pm=\frac{\partial \mathcal{W}}{\partial \lapm},\lapm\right) = E . 
\end{equation}
Trying a solution of the form $\mathcal{W} = W(\lap) + W(\lam)$, one finds that we need
\beq
\tilde{H} \left(\frac{\partial W}{\partial \la},\la\right) = (\alpha^2-\beta^2) E\la + V
\eeq
which is solved if
\begin{eqnarray}
\left(\frac{\partial W}{\partial \la}\right)^2  = \frac{\left\{V - \sum_a \prod_{b\neq a} (\omega_a^2-\omega_b^2)
\frac{C_a^2}{\la-\omega_a^2} + \left[(\alpha^2 - \beta^2)E + \alpha^2 \sum_a\omega_a^2\right] 
  \la-\alpha^2 \la^2\right\}}{\prod_a(\la-\omega_a^2)}.\nonumber
\end{eqnarray}
Here $V$ is a constant of motion related to $\tilde{H}$. Thus, the solution to the Hamilton-Jacobi equation is 
\begin{eqnarray} 
\mathcal{W}(\lapm,V,E) = W(\lap,V,E) + W(\lam,V,E)\ .  
\end{eqnarray}
Consequently, the equations of motion reduce to
\begin{eqnarray}
\frac{\partial W(\lap,V,E)}{\partial V} +  \frac{\partial W(\lam,V,E)}{\partial V} &=& U   \\
\frac{\partial W(\lap,V,E)}{\partial E} +  \frac{\partial W(\lam,V,E)}{\partial E} &=& \xi,  
\end{eqnarray}
where $U$ is a constant of integration.  Integrating these equations of motion, we obtain  
\begin{eqnarray}
\int^{\lap} \frac{d\la}{\sqrt{P_5(\la)}} + \int^{\lam} \frac{d\la}{\sqrt{P_5(\la)}} & = & 2U                                       \label{eom1_z} \\
\int^{\lap} \frac{\la\, d\la}{\sqrt{P_5(\la)}} + \int^{\lam} \frac{\la\, d\la}{\sqrt{P_5(\la)}} &=& \frac{2\xi}{\alpha^2-\beta^2} \label{eom2_z}
\end{eqnarray}
where $P_5(\la)$ is a quintic polynomial. The constraint (\ref{rcc1}) give $\mathcal{H} = E = 0$, in which case $P_5(\la)$ reduces to 
\begin{equation}
P_5(\la)=\prod_a(\la-\omega_a^2) \bigg\{V - \sum_a \prod_{b\neq a}(\omega_a^2-\omega_b^2)\frac{C_a^2}{\la-\omega_a^2} +\alpha^2 \la\sum_a\omega_a^2 
- \alpha^2 \la^2 \bigg\} 
\label{P5}
\end{equation}
Instead of using the Hamilton-Jacobi method, the same equation of motion can be derived by noting that both, the energy $H$ and 
\beq
 V = \frac{\tilde{H}(p_+,\lap) - \tilde{H}(p_-,\lam)}{(\alpha^2-\beta^2)(\lap-\lam)}
\eeq
are conserved for any Hamiltonian of the type (\ref{ztHam}). This can be verified by computing the Poisson bracket $\{H,V\}_{\mbox{P.B.}} =0$.  
Two conservation laws allows us to compute $\lap'$, $\lam'$ in terms of $\lap$, $\lam$, with the result (\ref{eom1_z}), (\ref{eom2_z}).

\subsection{Analysis of the solutions}

\begin{figure}
\epsfig{file=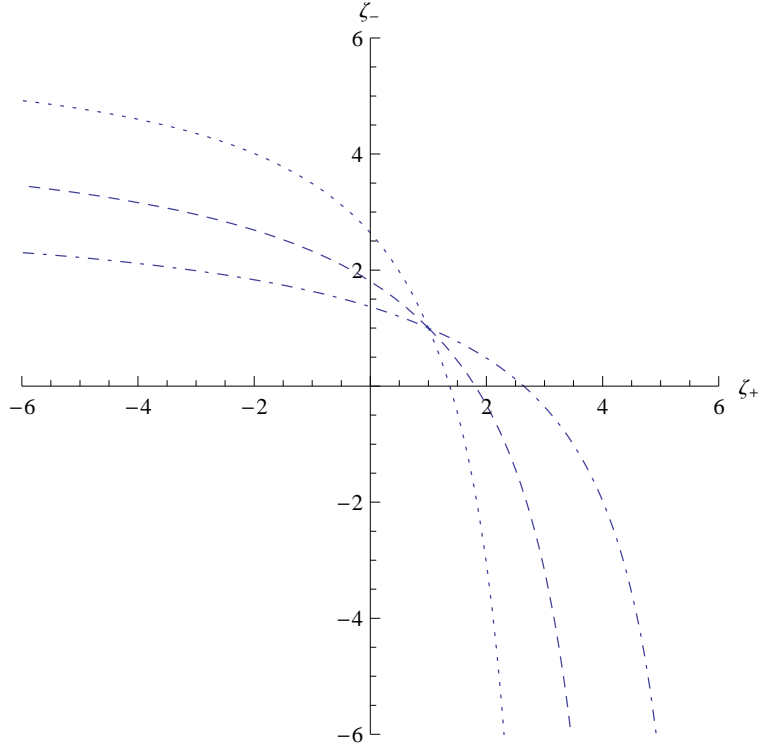, width=10cm}
\caption{Solutions for $\omega_1=3.5$, $\omega_2=1$, $\omega_3 = 4$, $C_1=2$, $\alpha=1$, $\beta=0.5$. Notice that all solutions 
cross at $\zeta_+=\zeta_-=\omega_2=1$. 
\label{fig:solution}}
\end{figure}

 All the dynamical information about the system is contained in the position of the roots of the polynomial $P_5$. The motion takes places
in regions where $P_5$ is positive. In our case we are interested in strings which reach the boundary, namely such that $r_a\rightarrow\infty$
at the ends of the string. From (\ref{r(zeta)}), we see that this corresponds to the region where one of the $\zeta$'s goes to infinity. 
In fact, one end of the string corresponds to $\zeta_+\rightarrow -\infty$ and the other to $\zeta_-\rightarrow -\infty$. 
The region $\la_\pm \rightarrow \infty$ is forbidden since there $P_5$ is negative under the radical (\ref{eom1_z},\ref{eom2_z}). 
In the appendix, we analyze the various possible ranges of variation for the frequencies $\omega_a$ and the variables 
$\zeta_\pm$ and conclude that the appropriate cases are  
\begin{eqnarray}
\textrm{Case 1:}\,\,\, \omega_3^2 > \omega_1^2 \geq \lam \geq \omega_2^2 \geq \lap  \\
\textrm{Case 2:}\,\,\, \omega_3^2 > \omega_1^2 \geq \lap \geq \omega_2^2 \geq \lam . 
\label{cases}
\end{eqnarray}
 These two cases correspond to two branches of the solution describing the two ends of the string. The two branches meet at $\zeta_+=\zeta_-=\omega_2^2$. 
We can see that this requires that $\omega_2$ is a root of $P_5$. Furthermore, we need $P_5$ to be positive on both sides of $\omega_2$ which means that
$\omega_2$ is a double root. This requires
\begin{equation}
C_2 = 0 \ , \ C_1 \omega_1 + C_3 \omega_3 =0 \label{C1C3}
\end{equation}
and
\begin{equation}
V = \frac{C_1^2(\omega_1^2 - \omega_3^2)^2 - \alpha^2 \omega_2^2\omega_3^2(\omega_1^2 + \omega_3^2)}{\omega_3^2}.
\end{equation}
Finally, with the value of V fixed, no other freedom remains in the form of $P_5$ and the remaining three roots are determined to be
\begin{eqnarray}
\lambda_0 &=& \omega_1^2 + \omega_3^2 \\
\lambda_+ &=& \frac{\omega_1^2 + \omega_3^2}{2} + \frac{\sqrt{1 + 4\frac{C_1^2}{\alpha^2\omega_3^2}}\,(\omega_3^2 - \omega_1^2)}{2}  \\
\lambda_- &=& \frac{\omega_1^2 + \omega_3^2}{2} - \frac{\sqrt{1 + 4\frac{C_1^2}{\alpha^2\omega_3^2}}\,(\omega_3^2 - \omega_1^2)}{2}
\end{eqnarray}
which yields 
\begin{equation}
P_5(\zeta) = -\alpha^2 (\zeta - \omega_2^2)^2(\zeta - \lambda_0)(\zeta - \lambda_+ )(\zeta - \lambda_-).
\label{poly}
\end{equation}
Notice that we need $\omega_2^2 <\lam < \omega_1^2$ which imposes a restriction on the values of $C_1$ that we can choose. 
Examples of solutions to the equations of motion (\ref{eom1_z}), (\ref{eom2_z}), obtained numerically, are given in fig.\ref{fig:solution}. 
The solutions have two branches, one such that $-\infty<\zeta_-<\omega_2^2<\zeta_+<\bar{\zeta}_+$ and the other such that 
$-\infty<\zeta_+<\omega_2^2<\zeta_-<\bar{\zeta}_-$. 
One of the constants $\bar{\zeta}_\pm$ can be chosen arbitrarily in the interval $(\omega_2^2,\omega_1^2)$ but the other should then be chosen such 
that the slopes of both branches match at $\zeta_\pm=\omega_2^2$. In fact, in the figure the different solutions have equal values of $\omega_a$ 
and $C_1$ and differ only on $\bar{\zeta}_+$. 

For the polynomial (\ref{poly}), the integrals in eqs.(\ref{eom1_z}) and (\ref{eom2_z}) can be computed in terms of elliptic integrals. 
It is also illuminating to analyze the asymptotic behavior near the boundary, namely when one of the $\zeta$'s goes to infinity. 
 Differentiating the equations of motion, (\ref{eom1_z}) and (\ref{eom2_z}), with respect to $\xi$ yields,
\begin{eqnarray}
\frac{\lap'}{\sqrt{P_5(\lap)}} + \frac{\lam'}{\sqrt{P_5(\lam)}} & = & 0                         \\
\frac{\lap'\lap}{\sqrt{P_5(\lap)}} + \frac{\lam'\lam}{\sqrt{P_5(\lam)}} &=& \frac{2}{\alpha^2-\beta^2},
\end{eqnarray}
or, equivalently,
\begin{eqnarray}
\lap' = \pm\frac{2}{\alpha^2-\beta^2}\frac{\sqrt{P_5(\lap)}}{(\lap - \lam)}  \\
\lam' = \pm\frac{2}{\alpha^2-\beta^2}\frac{\sqrt{P_5(\lam)}}{(\lam - \lap)} ,
\end{eqnarray}
where we emphasize that we can choose (independently) both signs of the square root. 
Consider the limit $\lam(\xi) \rightarrow -\infty$, $\lap(\xi) \rightarrow \bar{\lap}$ where $\bar{\lap}$ is a constant. 
In that limit, the equations of motion can be rewritten as follows,
\begin{eqnarray}
\lap' &\simeq& \pm\frac{2}{\alpha^2-\beta^2}\frac{\sqrt{P(\bar{\zeta}_+)}}{(-\lam)}\\
\lam' &\simeq& \pm\frac{2\alpha}{\alpha^2-\beta^2}(-\lam)^{\frac{3}{2}} ,
\end{eqnarray}
which gives 
\begin{eqnarray}
\lam(\xi) \simeq -\frac{(\alpha^2-\beta^2)^2}{\alpha^2} \frac{1}{(\xim-\xi)^2}, \ \ \ \ \ (\xi\rightarrow\xim) .
\label{asympt}
\end{eqnarray}

\subsection{Energy and Angular Momentum} 

 We now proceed to compute the energy and angular momentum of the solutions. First, however, we have to discuss the issue of the boundary 
conditions we use at the end points of the string.  For open strings, the momentum is conserved when Neumann boundary conditions are imposed 
at the end points. This boundary condition also ensures that such momentum can be computed as an integral over any path on the worldsheet 
that goes from one boundary to the other. From a physical point of view the b.c. ensures that there is no momentum flow out or into the string 
at the end points. 

In our case, and in general when doing Wilson loop computations, one imposes Dirichlet boundary conditions at the end points, namely when the 
string reaches the boundary. For that reason, in general there is a momentum (and energy) flow into the string. If the string is rigid, namely, 
its shape does not change in time, then the momentum flow at one end is compensated by the one at the other end and the total momentum is conserved. 

Although conserved, the definition of the total momentum now depends on the end points of the integration path taken. We choose to define 
it through a path that reaches both end points of the string at the same global time. In static gauge where one chooses $t=\tau$,  
a path of constant $\tau$ has 
such property. In conformal gauge however, if we take a constant $\tau$ path we need to add an extra leg at the boundary as shown in 
fig.(\ref{fig:paths}). As shown below, the integral over path (c) in the figure is easily done by noticing that it is a path of constant $\xi$. 

\begin{figure}
\epsfig{file=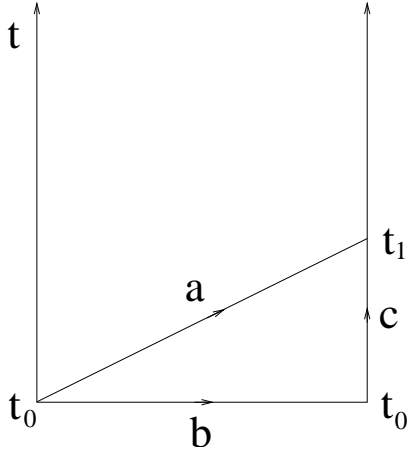, height=6cm}
\caption{Different paths one can use to compute the energy and momentum. We use path (b) as our definition. It can then be computed by subtracting
paths (a) and (c). Since we use Dirichlet boundary condition path (c) does not vanish as is the case with Neumann b.c. 
\label{fig:paths}}
\end{figure}

 If the coordinates $X_a$ in the Lagrangian (\ref{CGL}) are parameterized as $X_a=r_ae^{i\phi_a}$, it can be seen that the momenta conjugate to the angles
$\phi_a$ are conserved. The momentum conjugated to $\phi_3$ is the energy (since $\phi_3$ is the global time $t$) whereas the momenta conjugated to 
$\phi_{1,2}$ are angular momenta, denoted as $\J_{1,2}$. The corresponding conserved currents are denoted as
\beq
\J_{a}^{\sigma} = -\eta_a r_a^2 \phi_a{}', \ \ \ \J_{a}^{\tau} = \eta_a r_a^2 \dot{\phi}_a{}
\eeq
With the conserved current we proceed to integrate over path (b) in fig.(\ref{fig:paths}) or alternatively subtracting paths (a) and (c):
\begin{eqnarray}
\mathcal{P}_a &=& \left\{ \int_{(a)} -\int_{(c)}\right\}\ \left[ \J^\tau_a d\sigma - \J^\sigma_a d\tau\right] \\
              &=&   \frac{\beta C_a \Deltaxi}{\alpha (\alpha^2-\beta^2)} 
                  + \frac{\eta_a \alpha\omega_a}{\alpha^2-\beta^2}\int d\xi\, r_a^2 - \frac{C_a}{\alpha \omega_3} \Delta \mu_3,
\end{eqnarray}
where we used that $\phi_a = \mu_a+ \omega_a \tau $, namely $\dot{\phi}_a = \beta \mu'_a+\omega_a$.  Also the integral over path (c) is easily 
done noticing that $\J_a^{\sigma,\tau}$ depend only on $\xi$ which is constant along (c). Therefore
\beq
\int_{(c)} \left[ \J^\tau_a d\sigma - \J^\sigma_a d\tau\right] = 
-\frac{1}{\alpha} \left.\left[\beta \J_a^\tau + \alpha \J_a^\sigma\right]\right|_{\xi=\xi_{max}} \int_{(c)} d\tau =\frac{C_a}{\alpha \omega_3} \Delta \mu_3
\eeq
where $\Delta \mu_3 = \mu_3(\xi_{max})-\mu_3(\xi_{min})$, namely the difference in $\mu_3$ between the end points of the string. It will be computed in 
the next subsection. Using the conformal constraint (\ref{rcc2}) and that $C_2 = 0$ from the form of $P_5$, the energy and two angular momenta can 
be derived as:
\begin{eqnarray}
\mathcal{E}   &=& 
   T \frac{\beta C_3\Deltaxi}{\alpha(\alpha^2-\beta^2)}
  - T \frac{\alpha \omega_3}{\alpha^2-\beta^2}\int_{\xi_{min}}^{\xi_{max}}\!\!\!\!\! d\xi\,r_3^2 - \frac{C_3}{\alpha \omega_3} \Delta \mu_3,
  \nonumber \\
\J_1 &=&  T \frac{\beta C_1\Deltaxi}{\alpha (\alpha^2-\beta^2)} 
        + T \frac{\alpha\omega_1}{\alpha^2-\beta^2}\int_{\xi_{min}}^{\xi_{max}}\!\!\!\!\! d\xi\, r_1^2 - \frac{C_1}{\alpha \omega_3} \Delta \mu_3,
    \label{charges} \\
\J_2 &=& T \frac{\alpha\omega_2}{\alpha^2-\beta^2}\int_{\xi_{min}}^{\xi_{max}}\!\!\!\!\! d\xi\, r_2^2 .\nonumber
\end{eqnarray}
where we restored the string tension $T$ and defined $\Delta \xi=\xi_{max}-\xi_{min}$. The constant $C_3$ can be eliminated using $\omega_1C_1+\omega_3C_3=0$ according to eq.(\ref{C1C3}). 
 For strings ending on the boundary, the integrals have a divergence that should be subtracted as we describe below. Indeed,
\begin{equation}
\int d\xi\, r_a^2(\xi)= -\eta_a\int d\xi \frac{(\lap-\omega_a^2)(\lam-\omega_a^2)}{\prod_{b\neq a} (\omega_a^2-\omega_b^2)} ,
\end{equation}
generically diverge as can be seen using the asymptotic behavior derived in (\ref{asympt}):
\begin{equation}
\int d\xi\, r_a^2(\xi) \approx \eta_a\int^{\xim} d\xi\, \frac{(\bar{\lap}-\omega_a^2)}{\prod_{b\neq a} 
   (\omega_a^2-\omega_b^2)}\Big(\frac{\alpha^2-\beta^2}{\xim-\xi}\Big)^2 .
\end{equation}
 To extract the leading behavior of the divergence, let us perform the integrations up to a value $\bar{\xi} \lesssim \xim$.  Using this $\bar{\xi}$, 
define a radius $R= r_3(\bar{\xi})$ to characterize the radial extension of the string. We have
\begin{equation}
r_3^2(\bar\xi) = 
      R^2 \simeq \frac{(\omega_3^2-\bar{\lap})}{\prod_{b\neq 3} (\omega_3^2-\omega_b^2)}\Big(\frac{\alpha^2-\beta^2}{\xim-\bar{\xi}}\Big)^2
\end{equation}
or conversely
\begin{equation}
(\xim-\bar{\xi})^2 = \frac{(\omega_3^2-\bar{\lap})}{\prod_{b\neq 3} (\omega_3^2-\omega_b^2)}\frac{(\alpha^2-\beta^2)^2}{R^2} .
\end{equation}
Finally, the divergent piece of the integrals is
\begin{eqnarray}
\int d\xi\, r_a^2(\xi) &\approx& \lim_{\bar{\xi}\rightarrow \xim}\,\frac{\eta_a(\bar{\lap}-\omega_a^2)}{\prod_{b\neq a} 
            (\omega_a^2-\omega_b^2)}\frac{\alpha^2-\beta^2}{\xim- \bar{\xi}}\\
&=& \lim_{R\rightarrow\infty}\,(\alpha^2-\beta^2) \frac{\eta_a (\bar{\lap}-\omega_a^2)}{\prod_{b\neq a} (\omega_a^2-\omega_b^2)}
                            \sqrt{\frac{\prod_{b\neq 3}(\omega_3^2-\omega_b^2)}{\omega_3^2-\bar{\lap}}} R
\end{eqnarray}
namely the energy and momenta diverge linearly in $R$. This is actually well-known from the Wilson loop computations in \cite{WL1}. The only
difference is that now the end point of the string is moving and the result is essentially a boost of the static string. To make this explicit,
we should compute the velocity, on the boundary, of the end point of the string. In global coordinates (\ref{global}), the asymptotic
value of $\theta$ is a constant that we define as $\theta_0=\theta(\xim)$. In the limit $r_a\rightarrow\infty$, 
$\zeta_-\rightarrow\infty$ and $\lap\rightarrow\bar{\zeta}_+$, we get
\beqa
\lim_{\rho\rightarrow\infty}\frac{r^2_1}{r^2_3}= \sin^2\theta_0\
  &=&\frac{\bar{\zeta}_+-\omega_1^2}{\bar{\zeta}_+-\omega_3^2}\frac{\omega_3^2-\omega_2^2}{\omega_1^2-\omega_2^2}, \\
\lim_{\rho\rightarrow\infty}\frac{r^2_2}{r^2_3}= \cos^2\theta_0\
  &=&\frac{\bar{\zeta}_+-\omega_2^2}{\bar{\zeta}_+-\omega_3^2}\frac{\omega_3^2-\omega_1^2}{\omega_2^2-\omega_1^2} .
\label{theta0}
\eeqa
 The metric in the boundary is
\beq
ds^2 = -dt^2 + d\theta^2+ \sin^2\theta d\phi_1^2 + \cos^2\theta d\phi_2^2 .
\eeq
Consequently, the velocity in the angular directions is given by
\begin{eqnarray}
v_1^2 &=& \sin^2\theta_0\Big(\frac{\omega_1}{\omega_3}\Big)^2\\
v_2^2 &=& \cos^2\theta_0\Big(\frac{\omega_2}{\omega_3}\Big)^2
\end{eqnarray}
and 
\beq
v^2   = v_1^2+v_2^2 = \sin^2{(\theta_0)}\Big(\frac{\omega_1}{\omega_3}\Big)^2 + \cos^2{(\theta_0)}\Big(\frac{\omega_2}{\omega_3}\Big)^2 
\eeq
 Using eq.(\ref{theta0}) we can compute $v_{1,2}$ in terms of $\omega_a$ and $\bar{\zeta}_+$. In particular, it is useful to notice that
\beq
1-v^2 = \frac{(\omega_3^2-\omega_1^2)(\omega_3^2-\omega_2^2)}{\omega_3^2(\omega_3^2-\bar{\lap})}
\eeq
which shows that $v<1$, namely the particles in the boundary cannot move at the speed of light within this ansatz. It appears that if 
$\omega_1=\omega_3$ or $\omega_1=\omega_2$ then $v=1$ but when two $\omega$'s coincide the ansatz we are using is not valid. The case of equal
frequencies will be analyzed in a later section.  
 With these results we can rewrite the divergent part of the energy and angular momenta as 
\beqa
 E &\simeq&  \frac{T}{\sqrt{1-v^2}}\ R \\
\J_1 &\simeq& T\, \frac{v_1\sin\theta_0}{\sqrt{1-v^2}}\   R  \label{divJ1} \\
\J_2 &\simeq& T\, \frac{v_2 \cos\theta_0}{\sqrt{1-v^2}}\  R 
\eeqa
 In the appendix, the divergence for a straight string moving in direction $x$ with velocity $v$ is derived as
\beq
E \simeq \frac{T}{\sqrt{1-v^2}} \frac{1}{\epsilon} , \ \ \ P_x\simeq T\, \frac{v}{\sqrt{1-v^2}} \frac{1}{\epsilon}  .
\eeq
 If, for example, we identify $x=\phi_1 \sin\theta_0 $ then we have $\J_1 = P_x \sin\theta_0$ and the divergence agrees with eq.(\ref{divJ1}). 
The same is true for $\J_2$ and $\E$. We find then that the divergencies agree with the usual one and should be subtracted. 
 For brevity, we still use the formulas (\ref{charges}) with the understanding that the diverge is subtracted.  Unfortunately the integrals in 
(\ref{charges}) cannot be done in terms of known functions. It is easy however to obtain numerical values for specific solutions if so is desired. 
 

\subsection{Angular Separation} 

The string rotates rigidly, namely its shape does not change in time. In particular, the angular separations between the
end points is constant in time. To compute them, we use global coordinates (\ref{global}), through the following identifications:
\beq
\phi_1 = \mu_1 + \omega_1 \tau, \ \ \ 
\phi_2 = \mu_2 + \omega_2 \tau, \ \ \
t      = \mu_3 + \omega_3 \tau \label{t} .
\eeq
At a given constant time  $t$, chosen to be zero ($t=0$), eq.(\ref{t}) gives, for $\tau$,
\begin{equation}
\tau = -\frac{1}{\omega_3}\mu_3 .
\end{equation}
Consequently,
\beq
\phi_1 = \mu_1 - \frac{\omega_1}{\omega_3} \mu_3 \label{phi1}, \ \ \ 
\phi_2 =  \mu_2 - \frac{\omega_1}{\omega_3} \mu_3 \label{phi2} ,
\eeq
and, finally
\begin{eqnarray}
\Delta \phi_1 &=& \int^{\xim}_{\xi_{\mathrm{min}}} \left[ \mu_1' - \frac{\omega_1}{\omega_3} \mu_3'\right] d\xi\label{delphi1} \\
\Delta \phi_2 &=& \pi+ \int^{\xim}_{\xi_{\mathrm{min}}} \left[ \mu_2' - \frac{\omega_1}{\omega_3} \mu_3'\right] d\xi \label{delphi2}
\end{eqnarray}
 where we assumed that the two branches already mentioned match at $\xi=0$ and extend from $(\xi_{\mathrm{min}},0)$ and $(0,\xim)$ respectively. 
There is an extra jump in $\pi$ on $\phi_2$ since $r_2$ becomes zero at $\xi=0$. When the trajectory crosses the origin in the plane $(\r_2,\mu_2)$ 
we need to add $\pi$ to the angle $\mu_2$. Although the evaluation of the integrals seems to require the explicit form of $r_a(\xi)$,
this can be circumvented by converting the $\xi$ integrals into integrals over $\lapm$. This can be accomplished by expressing the radial variables 
in terms of $\lapm$, through (\ref{r(zeta)}) and then using the equations of motion for $\lapm$, \ie\ eqs. (\ref{eom1_z}) and (\ref{eom2_z}).
The result is
\begin{eqnarray}
\Delta \phi_1 &=& \frac{C_1}{2} (\omega_1^2-\omega_3^2) \left\{ 
                  \left[\int_{-\infty}^{\bar{\zeta}_+}+\int_{-\infty}^{\bar{\zeta}_-}\right]
                  \frac{d\zeta}{(\zeta-\omega_2^2) \sqrt{P_3(\zeta)}} 
                  \left(\frac{\omega_1^2-\omega_2^2}{\zeta-\omega_1^2}-\frac{\omega_1^2}{\omega_3^2}\frac{\omega_3^2-\omega_2^2}{\zeta-\omega_3^2}\right)
                  \right\} \nonumber \\
\Delta \phi_2 &=& \frac{\omega_1\omega_2}{\omega_3^2}\frac{C_1}{2} (\omega_3^2-\omega_1^2)(\omega_3^2-\omega_2^2) \left\{ 
                  \left[\int_{-\infty}^{\bar{\zeta}_+}+\int_{-\infty}^{\bar{\zeta}_-}\right]
                  \frac{d\zeta}{(\zeta-\omega_3^2)(\zeta-\omega_2^2)\sqrt{P_3(\zeta)}} 
                  \right\}
\end{eqnarray}
where $P_3(\zeta)=-\alpha^2(\zeta-\lambda_0)(\zeta-\lambda_+)(\zeta-\lambda_-)$, namely $P_5(\zeta)=(\zeta-\omega_2^2)^2 P_3(\zeta)$. 
The integrals go over the pole at $\zeta=\omega_2^2$ and should be understood in the principal part sense. They can be expressed in terms 
of elliptic integrals ($\Pi$, $F$) by defining
\beq
G_j^\pm= \int_{-\infty}^{\bar{\zeta}_\pm} \frac{d\zeta}{(\zeta-\omega_j^2)\sqrt{P_3(\zeta)}} = \frac{2}{(\omega_j^2-a)\sqrt{a-c}} 
        \left[\Pi\left(\alpha_{\pm},\frac{a-\omega_j^2}{a-c},p\right)-F(\alpha_\pm,p)\right] 
\eeq
where
\beq
\sin \alpha_\pm= \sqrt{\frac{a-c}{a-\bar{\zeta}_\pm}}, \ \ \ \ p=\sqrt{\frac{a-b}{a-c}} \label{alphapdef}
\eeq
and $c<b<a$ are the roots of $P_3(\zeta)$ ordered from smaller to larger. That is, we need to order $\lambda_0$, $\lambda_\pm$ accordingly 
(the actual order depends on the value of the parameters). The elliptic integrals $F$, and $\Pi$ are as defined in \cite{GR}. Thus, we obtain
\beqa
\Delta\phi_1 &=& \frac{(\omega_1^2-\omega_3^2)C_1}{2\omega_3^2}  \left\{\omega_3^2\left(G_1^++G_1^-\right)
                     -\omega_1^2\left(G_3^++G_3^-\right)
                     +(\omega_1^2-\omega_3^2)\left(G_2^++G_2^-\right)\right\} \nonumber\\
\Delta\phi_2 &=& \frac{\omega_2 (\omega_1^2-\omega_3^2)C_3}{2\omega_3}  \left\{G_3^++G_3^--G_2^+-G_2^-\right\}
\eeqa
Moreover, the difference $\Delta \mu_3$ which appears in the computation of the 
energy and angular momenta (see eq. (\ref{charges})) can be obtained as:
\beq
\Delta\mu_3 = \beta \omega_3 \Delta \xi + \frac{C_3}{2\omega_3^2}(\omega_3^2-\omega_1^2)
                \left[\omega_3^2\left(G_3^++G_3^-\right)
                      -(\omega_3^2-\omega_2^2)\left(G_2^++G_2^-\right)\right]
\eeq
The difference $\Delta\xi=\xi_{max}-\xi_{min}$ which also appears in eq.(\ref{charges}) can be also evaluated in terms of elliptic integrals as:
\beq
\Delta\xi = F(\alpha_+,p) + F(\alpha_-,p) +\omega_2^2 \left(G_2^++G_2^-\right)
\eeq
where $\alpha_{\pm},p$ are the ones defined in eq.(\ref{alphapdef}). 

%

\section{Solution with $\omega_1 = \omega_3$}

The degenerate case when  two of the $\omega_a$'s coincide should be treated separately because in such case the change of 
variables (\ref{r(zeta)}) becomes singular. If $\omega_1=\omega_2$ then the system has an $SO(4)$ rotational symmetry which simplifies
the equations considerably. In this section we consider the, perhaps more interesting, case where $\omega_1=\omega_3$ 
(or equivalently $\omega_2=\omega_3$). In that case we have an enhanced symmetry to $SO(2,2)$ which also helps simplifying the problem.
Still the system has two angular momenta and behavior similar to the one we studied but in a somewhat simplified situation. 

The extra symmetry present when  $\omega_1=\omega_3$ can be made manifest with the change of variables
\begin{eqnarray}
x_1(\xi) &= z_1(\xi) + iz_2(\xi) =& z(\xi)\sinh{\psi(\xi)} \ e^{i\mu_1(\xi)} \nonumber \\
x_2(\xi) &= z_3(\xi) + iz_4(\xi) =& r_2(\xi)               \ e^{i\mu_2(\xi)} \nonumber \\
x_3(\xi) &= z_5(\xi) + iz_6(\xi) =& z(\xi)\cosh{\psi(\xi)} \ e^{i\mu_3(\xi)}. \label{psi_coord}
\end{eqnarray}


The hyperbolic constraint is
\begin{equation}
-1 = z_1^2 + z_2^2 + r_2^2 - z_5^2 - z_6^2 = -z^2 + r_2^2.
\end{equation}

The Lagrangian can be found by direct substitution of the change of variables into (\ref{Lag}) yielding,
\begin{eqnarray}
\mathcal{L} &=& (\alpha^2-\beta^2)\left[-z'^2 +z^2\psi'^2 + z^2\sinh^2{(\psi)}\mu_1'^2 
              - z^2\cosh^2{(\psi)}\mu_3'^2 +r_2'^2 + r_2^2\mu_2'^2\right] \nonumber \\
            &&  -2\beta\left[\omega_1(z^2\sinh^2{(\psi)}\mu_1' - z^2\cosh^2{(\psi)}\mu_3') + \omega_2r_2^2\mu_2'\right]  \nonumber \\
            &&   + \left[\omega_1^2z^2 -\omega_2^2r_2^2 \right] + \Lambda[1 + r_2^2 -z^2]. \,\,\,
\end{eqnarray}
The momenta canonically conjugate to each coordinate are
\begin{eqnarray}
P_z = \frac{\partial \mathcal{L}}{\partial z'}       &=& -2(\alpha^2-\beta^2)z'                                     \nonumber \\
P_2 = \frac{\partial \mathcal{L}}{\partial r'_2}     &=&  2(\alpha^2-\beta^2)r_2'                                   \nonumber \\
P_\psi = \frac{\partial \mathcal{L}}{\partial \psi'} &=&  2(\alpha^2-\beta^2)z^2\psi'                               \nonumber \\
J_1 = \frac{\partial \mathcal{L}}{\partial \mu'_1}  &=&  
          2(\alpha^2-\beta^2)z^2\sinh^2{(\psi)}\left[\mu'_1-\frac{\beta\omega_1}{(\alpha^2-\beta^2)}\right]\nonumber \\
J_2 = \frac{\partial \mathcal{L}}{\partial \mu'_2}  &=&  
          2(\alpha^2-\beta^2)r_2^2             \left[\mu'_2-\frac{\beta\omega_2}{(\alpha^2-\beta^2)}\right]\nonumber \\
J_3 = \frac{\partial \mathcal{L}}{\partial \mu'_3}  &=& 
         -2(\alpha^2-\beta^2)z^2\cosh^2{(\psi)}\left[\mu'_3-\frac{\beta\omega_1}{(\alpha^2-\beta^2)}\right]. 
\end{eqnarray}
Thus, the Hamiltonian is
\begin{eqnarray}
\mathcal{H} &=& \frac{1}{(\alpha^2-\beta^2)}\Bigg[\frac{1}{4}\left(-P_z^2 + \frac{P_\psi^2}{z^2} 
               + \frac{J_1^2}{z^2\sinh^2{(\psi)}} - \frac{J_3^2}{z^2\cosh^2{(\psi)}}{} +  P_2^2 + \frac{J_2^2}{r_2^2}\right) \nonumber \\
            &&  + \beta(\omega_1J_1 + \omega_2J_2 + \omega_1J_3) +\alpha^2(-\omega_1^2z^2 + \omega_2^2r_2^2)\Bigg].
\end{eqnarray}
The equations of motion associated with $\mu_{a=1,2,3}$ imply the conservation of the corresponding momenta $J_{a=1,2,3}$. 
The equation for $\psi$ is equivalent to the conservation of the total $SO(2,2)$ angular momentum $J^2$ defined as
\begin{equation}
J^2 = P_{\psi}^2 + \frac{J_1^2}{\sinh^2{\psi}} -\frac{J_3^2}{\cosh^2{\psi}}
\end{equation}
The remaining equation is equivalent to the conservation of the Hamiltonian.   
%
Similarly as in the general case, the two constraints can be written as:
\beqa
 \cH &=& 0 \\
 \omega_1J_1 + \omega_2J_2 + \omega_1J_3 &=& 0. \label{zcc2}
\eeqa
 The Hamiltonian in terms of the conserved momenta reads
\begin{eqnarray}
\mathcal{H} &=&  \frac{1}{\alpha^2-\beta^2}\left[\frac{1}{4}(-P_z^2+ \frac{J^2}{z^2} + P_{r_2}^2 + \frac{J_2^2}{r_2^2}) 
                + \beta(\omega_1(J_1 + J_3) + \omega_2 J_2) \right. \nonumber \\ 
             &&   \left. + \alpha^2(\omega_2^2 r_2^2 -\omega_1^2z^2)\right].
\end{eqnarray}

\subsection{Shape of the String}

 The angular motion is determined by the conservation of angular momenta. The other two variables $r_2$, $z$ are related by the constraint $r_2^2-z^2=-1$
reducing the system to a one-dimensional problem whose equation of motion, from energy conservation, is 
\begin{equation}
4(\alpha^2-\beta^2)^2z^2z'^2 = -\left[J_2^2z^2 + (z^2 -1)\Big(J^2 + 4\alpha^2z^2[(\omega_2^2 -\omega_1^2)z^2 -\omega_2^2]\Big)\right].
\end{equation}
which can be integrated to 
\begin{equation}
\int \frac{d(z^2)}{\sqrt{P_3(z^2)}} = \int \frac{d\xi}{\alpha^2 -\beta^2} = \frac{\xi - \xi_o}{\alpha^2-\beta^2}.
\end{equation}
where we defined the cubic polynomial
\beq
P_3(x) = 4\alpha^2(\omega_1^2 -\omega_2^2)x^3 + 4\alpha^2(2\omega_2^2 -\omega_1^2)x^2 - (J^2 +J_2^2 + 4\alpha^2\omega_2^2)x + J^2. 
 \label{P3}
\eeq
 Notice that $x = z^2 \geq 0$ and, furthermore, we need $P_3(x)\ge 0$.  Let $\lambda_1$, $\lambda_2$, and $\lambda_3$ be the roots of $P_3$, then 
the general form of the polynomial is
\begin{equation}
P_3(x) = 4\alpha^2(\omega_1^2 - \omega_2^2)(x-\lambda_1)(x-\lambda_2)(x-\lambda_3). \label{P3L}
\end{equation}
 If $\omega_2^2 > \omega_1^2$, the motion is bound (because $P_3(x\rightarrow\infty)\rightarrow-\infty$) whereas it is unbound in the opposite 
case $\omega_1^2 > \omega_2^2$. Here we are interested in the string reaching the boundary ($z\rightarrow\infty$) so we analyze the latter. Computing
$P_3(0)$ using eqs.(\ref{P3}) and (\ref{P3L}) we find
\beq
-\lambda_1\lambda_2\lambda_3    = \frac{J^2}{4\alpha^2(\omega_1^2 - \omega_2^2)} >0.
\eeq
 Since we need at least one real and positive root (which determines the smallest value of $z$ for the string), this requires that we have
two positive and one negative real roots which we order as $\lambda_1<0<\lambda_2<\lambda_3$. In terms of the roots, the integral for $z$ can be written as
\beq
 F(\mu,q) = \frac{\sqrt{2}}{ \sqrt{\alpha^2(\omega_1^2-\omega_2^2)(\lambda_3-\lambda_1)}} \frac{\xi-\xi_0}{\alpha^2-\beta^2}
\eeq
where $F(\mu,q)$ is a standard elliptic integral and
\beq
\mu = \arcsin\left(\sqrt{\frac{z^2-\lambda_3}{z^2-\lambda_2}}\right), \ \ \ \ q=\sqrt{\frac{\lambda_2-\lambda_1}{\lambda_3-\lambda_1}}
\eeq
 After obtaining an integral for $z(\xi)$, we can compute $\psi$ from its equation of motion which has the form
\begin{equation}
2(\alpha^2-\beta^2)z^2{\psi'} = \pm \sqrt{J^2 - \frac{J^2_1}{\sinh^2{(\psi)}} + \frac{J^2_3}{\cosh^2{(\psi)}}}.
\end{equation}
This can be simplified by defining $\Phi(\psi) = \cosh{(2\psi)}$ and parameterizing the string with a new variable $\zeta(\xi)$  such that
\begin{equation}
\frac{d\zeta}{d\xi} = \frac{1}{z^2(\xi)}.
\end{equation}
The $\psi$ equation reduces to
\begin{equation}
\frac{1}{2}(\alpha^2-\beta^2)\partial_{\zeta}\Phi = \sqrt{Q(\Phi)}
\end{equation}
where the quadratic polynomial $Q(\Phi)$ is given by
\begin{equation}
Q(\Phi) = \frac{J^2}{4}\Phi^2 + \frac{J_3^2 - J_1^2}{2}\Phi - \frac{J^2 + 2J_1^2 + 2J_3^2 }{4}.
\end{equation}
Integrating, we find
\begin{equation}
\frac{\zeta - \zeta_o}{\alpha^2-\beta^2} = \frac{2}{J}\log{\left[\frac{J\sqrt{Q(\Phi)} 
  + J^2\Phi + J_3^2 - J_1^2}{\sqrt{-J^2(J^2 + 2J_1^2 + 2J_3^2) - (J_3^2 -J_1^2)^2}}\right]}
\label{Phisol}
\end{equation}
To understand the result, remember that, as before, $-\xi_0<\xi<\xi_0$ and the string reaches the boundary at the end points of the interval. 
From the equation of motion for $z$, we can derive that close to the boundary 
\beq
z \simeq 2\frac{\alpha^2-\beta^2}{|\xi-\xi_0|}, \ \ \ \ \ \xi\rightarrow \pm\xi_0.
\eeq
Recalling the relation between $\zeta$ and $\xi$, in this limit,
\begin{equation}
\int d\zeta = 4(\alpha^2-\beta^2)^2 \int d\xi(\xi_0 - \xi)^2 = \frac{4(\alpha^2-\beta^2)^2(\xi_0 - \xi)^3}{3}.
\end{equation}
The integral is finite, and as a consequence $\zeta$ also spans a finite interval and from eq.(\ref{Phisol}) also does $\Phi$ (and then $\psi$). 
Namely, $\psi(\xi_0)=\psi_0$ for some $\psi_0$. 

Our interest in the case $\omega_1=\omega_3$ was due to the possibility that the end points of the string could move at the speed of light.
However, in this case, the velocity of the end-points, at the boundary, is
\beq
 v^2 = \lim_{\xi\rightarrow \xi_0} \frac{\omega_1^2r_1^2+\omega_2^2r_2^2}{\omega_3^3r_3^2} 
     = 1 - \frac{\omega_1^2-\omega_2^2}{\omega_1^2} \frac{1}{\cosh^2 \psi_0}
\eeq
where we used the parameterization (\ref{psi_coord}) and the relation $\omega_1=\omega_3$.  Since we argued that $\psi_0$ is finite, we always have
$v^2$ strictly smaller than one. Again one can think of putting $\omega_1=\omega_2$ to get $v=1$. In such case $\omega_1=\omega_2=\omega_3$. However, all 
solutions with the three $\omega$'s equal can be converted to the simple one spin solution by means of an $SO(4,2)$ rotation in \ads{5} \cite{TT1}\footnote{We are grateful to A. Tirziu and A. Tseytlin for pointing this out.}. Therefore there are no new solutions with the particles moving at the speed of light.

\section{Conclusions}

 In this paper we considered Wilson loops with the shape of a double helix in space time. This corresponds to two particles rotating in an $S^3$.
When the system has only one angular momenta the solutions are simple and the resulting energy as a function of angular momenta can be thought
as an analytic continuation of the rotating string of \cite{GKP, spikes}. In fact they both go to the same limiting shape when the particles move at the 
speed of light.  In the case of two angular momenta we need to resort to the techniques of \cite{NR, DF} involving the integrable Neumann-Rosochatius 
system to find the solution. The result can be written in terms of (1-dimensional) integrals which can be evaluated numerically. We plotted some solutions
to illustrate the results. The conserved charges, namely energy and angular momenta are divergent but we show that the divergent piece, as expected, 
is canceled if we subtract the same quantities computed for a straight string moving with the same speed. Finally, in this case, there is no new limit 
in which the particles move at the speed of light, namely other than the case of only one non-vanishing angular momentum. 
Besides the new solutions a slight difference with more standard calculations is that we are interested in the energy and angular momentum of 
the Wilson loop rather then in its expectation value (which would be given by the area of the world-sheet). Similarly, from the field theory 
point of view we are interested in the energy and 
angular momentum of a quark and anti-quark which move on an $S^3$ in a prescribed way. We started to briefly analyze this configuration by 
considering the classical electromagnetic field produced by two charges moving in a sphere. A simple solution was found for the case where 
they move at the speed of light. It is interesting that the solution is regular, namely not a shock wave. It seems complicated
to extend this calculations to higher loops in the field theory. However, for the case of a closed string moving in the interior of \ads{} the dual
description in terms of operators in the SL(2) sector is well understood in terms of spin chains. Since, form the bulk point of view, the results are
related we expect that a similar description based on something analogous to a spin chain also exists for these Wilson loops. This should be an 
interesting topic for further research.

\begin{acknowledgments}
We are grateful to A. Tirziu for several comments and suggestions and to A. Tseytlin for discussions and collaboration on a 
closely related topic. This work was supported in part by NSF under grant PHY-0805948, by DOE under grant DE-FG02-91ER40681 
and by the Alfred P. Sloan Foundation. The work of A.I. was supported in part by a Lee Grodzins summer research grant (in honor of Anna Akeley).
\end{acknowledgments}

\section{Appendix} 

\subsection{Moving Straight String} 

 Since we are considering Wilson loops where the end point at the boundary moves in time it is useful to study the simplest possible case to 
check the divergences near the boundary. Such divergences should all be the same, depending only on the speed of the string.
Thus, consider \ads{3} space in Poincare coordinates $ds^2 = \frac{1}{z^2}\left(-dt^2+dx^2+dz^2\right)$ and a static string, of tension $T$, 
stretching down form the boundary at $z=0$ to the horizon at $z=\infty$. If we boost that string we obtain a solution such that
\beq
t = \tau, \ \ \ x = v\tau, \ \ \ z = \sigma. 
\eeq
 The energy of such string can be calculated as
\beq
P_0 =  \frac{T}{\sqrt{1-v^2}}\int_\epsilon^{\infty} d\sigma \frac{1}{z^2} 
    =  \frac{T}{2\pi\alpha'\sqrt{1-v^2}}\frac{1}{\epsilon}
\eeq
We see that the usual $\frac{1}{\epsilon}$ UV divergence gets multiplied by a Lorentz factor $\frac{1}{\sqrt{1-v^2}}$  as we also found in the 
more involved situation studied in the main text. Similarly the momentum diverges as
\beq
 P_x = T \frac{v}{\sqrt{1-v^2}} \frac{1}{\epsilon}
\eeq
which is useful to understand the divergence of the angular momenta for the rotating strings.

\subsection{Relative Magnitudes of $\omega_1$, $\omega_2$, $\omega_3$, and $\lapm$} 

From eq.(\ref{r(zeta)}), the three radial variables can be written in terms of the two unconstrained variables $\lapm$ as
\begin{eqnarray}
r_1^2 &=&   - \frac{(\lap - \omega_1^2)(\lam - \omega_1^2)}{(\omega_1^2 - \omega_2^2)(\omega_1^2 - \omega_3^2)} \label{r1} \\
r_2^2 &=&   - \frac{(\lap - \omega_2^2)(\lam - \omega_2^2)}{(\omega_2^2 - \omega_1^2)(\omega_2^2 - \omega_3^2)} \label{r2} \\
r_3^2 &=& \ \ \frac{(\lap - \omega_3^2)(\lam - \omega_3^2)}{(\omega_3^2 - \omega_1^2)(\omega_3^2 - \omega_2^2)} \label{r3} 
\end{eqnarray}
The fact that $r_a^2>0$ imposes some restrictions on the relative magnitudes of the $\omega_a$ and $\lapm$. Since there is a symmetry 
between $r_1$ and $r_2$ we can always choose $\omega_1>\omega_2$. Similarly, $\lapm$ enter equally in the previous equations so we can take 
$\lap>\lam$. Altogether we find three distinct possibilities
\begin{eqnarray}
 \lap &>& \omega_1^2 > \lam > \omega_2^2 > \omega_3^2 \label{res7} \\ 
 \lap &>& \omega_1^2 > \omega_3^2 > \omega_2^2 > \lam \label{res8} \\ 
\omega_3^2 &>& \omega_1^2 > \lap > \omega_2^2 > \lam \label{res9} 
\end{eqnarray}
Namely, given all possible orderings of the $\omega_a$'s (up to interchanging $\omega_{1,2}$) we choose the intervals where $\lapm$ can vary 
so that $r_a^2>0$. 

\subsection{Charges moving in a sphere} 

 From the field theory point of view the results of the paper refer to the strong coupling limit of the energy and angular momenta
of two charges of opposite sign moving on a 3-sphere. In this appendix we consider the situation from the perturbative point of view 
and compute the same results using the Maxwell equations, to which the non-abelian system reduces in the case of small coupling. 
The main result will be for charges moving in circles at the speed of light. 

Consider then the equation
\beq
D_\mu F^{\mu\nu} =0
\eeq
 where the covariant derivative $D_\mu$ refers to the metric of the sphere (and not to the non-abelian gauge field since we take $g_{YM}\rightarrow 0$). 
the metric is
\beq
ds^2 = -dt^2 +d\theta^2 +\sin^2\theta d\phi_1^2 +\cos^2\theta d\phi_2^2
\label{S3ma}
\eeq
 For the gauge field we make the ansatz that $A_{\phi_1}=0$, and that the other components are functions of $\theta$ and $\phi_2-\omega t$ 
as appropriate for the field produced by charges moving with angular velocity $\omega$ along the maximum circle $\theta=0$. 
Since the fields should be periodic in $\phi_2$ we further consider the Fourier modes:
\beqa
A_0^{(n)} &=& A^{(n)}_0(\theta)\ e^{i n(\phi_2-\omega t)} \\
A^{(n)}_\theta &=& A^{(n)}_\theta(\theta)\ e^{i n(\phi_2-\omega t)} \\
A^{(n)}_{\phi_2} &=& A^{(n)}_2(\theta)\ e^{i n(\phi_2-\omega t)}
\eeqa
 We can do a further gauge choice and eliminate one of the components. A convenient choice is to take $A_{\theta}=0$. In this way the equations simplify
and we obtain
\beq
 \partial_\theta A^{(n)}_0 = -\frac{1}{\omega\cos^2\theta} \partial_\theta A^{(n)}_2 \label{Arel}
\eeq
Finally, using that the functions should be regular at $\theta=\frac{\pi}{2}$ we find the unique solution
\beqa
 \partial_\theta A^{(n)}_2 &=& A_n y_n(\theta) \\
  y_n(\theta) &=& \frac{\scriptstyle \Gamma({ 1+\half n(1-\omega)})\Gamma({ 1+\half n(1+\omega)})}{\scriptstyle \Gamma({ 1+n})}
             F\left( { 1+\frac{1+\omega}{2} n, 1+\frac{1-\omega}{2} n; n +1;} \cos^2\theta\right) \sin\theta (\cos\theta)^{1+n} \nonumber
\label{ysol}
\eeqa
 where $F$ denotes the hypergeometric function and we took $n>0$. For $n<0$ we take $y_{-n}(\theta)=y_n(\theta)$. The coefficient 
was chosen such that $y_n(\theta) \simeq \frac{1}{\theta}$ when $\theta\rightarrow 0$.
The full solution is the superposition of the different Fourier modes taking into account that for $n=0$ 
there is no source since the total charge should be zero because the space is compact. The coefficients of the Fourier expansion can be computed
by matching with the field near the charges. In that region it should match the potential $A_0=\frac{q}{4\pi r}$ after an appropriate boost. 
For two charges $\pm q$ which at $t=0$ sit at $\phi_2=\pm\half \Delta \phi_2$ the solution reads
\beq
\partial_\theta A_2 = \frac{q\omega}{\pi^2} \sum_{n=1}^{\infty} y_n(\theta)  \sin\left(\frac{ n\Delta \phi}{2}\right) \sin\left(n(\phi_2-\omega t)\right)
\eeq
From here we can compute the electromagnetic field which gives
\beqa
 F_{\theta\phi_2} &=& \frac{q\omega}{\pi^2} \sum_{n=1}^{\infty} y_n(\theta)  \sin\left(\frac{ n\Delta \phi}{2}\right) \sin\left(n(\phi_2-\omega t)\right)\\
 F_{0\theta} &=& \frac{1}{\omega\cos^2\theta}  F_{\theta\phi_2} \\
 F_{0\phi_2} &=& \frac{q}{\pi^2} \frac{\cos\theta}{\sin\theta} \ \partial_\theta \sum_{n=1}^\infty \frac{1}{n}\frac{\sin\theta}{\cos\theta} y_n(\theta) 
     \sin\left(\frac{ n\Delta \phi}{2}\right) \cos\left(n(\phi_2-\omega t)\right)\label{Fres}
\eeqa
 where we used eq.(\ref{Arel}) to compute $F_{0\phi_2}$. Although this solves the problem of finding the electromagnetic field produced by 
two opposite charges moving along a maximum circle on an $S^3$ the result is not completely satisfactory because the expressions for the total 
energy and angular momentum are hard to evaluate. Nevertheless we present the calculation because in the particular limit $\omega\rightarrow 1$ 
the result simplifies and we obtain a particularly simple and interesting result. Indeed, for $\omega=1$ we have that the hypergeometric functions
in eq.(\ref{ysol}) can be evaluated in terms of elementary functions. The resulting series is a geometric series that can be summed with the result:
\beqa
 F_{\theta\phi_2} &=&  \frac{q}{2\pi^2}\frac{\cos^2\theta}{\sin\theta} 
\left[\frac{\cos\left(\xi-\half\Delta\phi\right)-\cos\theta}{1+\cos^2\theta-2\cos\theta\cos\left(\xi-\half\Delta\phi\right)}
 -\frac{\cos\left(\xi+\half\Delta\phi\right)-\cos\theta}{1+\cos^2\theta-2\cos\theta\cos\left(\xi+\half\Delta\phi\right)}\right] \nonumber \\
 F_{0\theta} &=& \frac{1}{\cos^2\theta}  F_{\theta\phi_2} \\ 
 F_{0\phi_2} &=& \frac{q}{2\pi^2} \left[ \frac{\cos\theta\sin\left(\xi-\half\Delta\phi\right)}
                   {1+\cos^2\theta-2\cos\theta\cos\left(\xi-\half\Delta\phi\right)}-
                   \frac{\cos\theta\sin\left(\xi+\half\Delta\phi\right)}
                   {1+\cos^2\theta-2\cos\theta\cos\left(\xi+\half\Delta\phi\right)}\right] \nonumber
\eeqa
which determines all the components of the electromagnetic field according to eq.(\ref{Fres}). It should be noted that the field are smooth
(except of course on top of the charge). This is in contrast to the case in flat space where a charge moving at the speed if light produces
a singular shock-wave. In fact near each charge (\eg\ $\xi-\half\Delta\phi\sim\theta^2\rightarrow 0$) the metric can be approximated by a 
pp-wave and the solution we found actually reduces to the field of a charge moving in a pp-wave found in \cite{KT}. For that reason, the 
divergence of the energy momentum tensor are exactly the same as in \cite{KT} which we already know reproduces the one-loop cusp anomaly.  
To match with the pp-wave in \cite{KT} one needs to identify $x_\pm=\frac{(\phi_2\pm t)}{\sqrt{2}}$, $r=\theta$, $x_1=\theta\cos\phi_1$, 
$x_2=\theta\sin\phi_1$ and $\mu=\frac{1}{\sqrt{2}}$ which follows from taking the pp-wave limit for the metric of $t\times S^3$, 
\ie\ eq.(\ref{S3ma}) and matching with the pp-wave metric in \cite{KT}.


\end{document}